\documentclass[letterpaper, 10 pt, conference]{ieeeconf}
\IEEEoverridecommandlockouts

\usepackage{times}

\usepackage{graphicx}
\usepackage{amsmath}
\usepackage{amsfonts}

\usepackage{longtable,tabularx}
\usepackage{mathtools}
\DeclarePairedDelimiter{\norm}{\lVert}{\rVert}

\usepackage{amssymb}

\newtheorem{rmk}{Remark}

\usepackage{multicol}

\usepackage{algorithm}
\usepackage{algpseudocode}

\usepackage{microtype}

\usepackage{cite}

\usepackage{xcolor}
\usepackage{xfrac}

\usepackage{hhline}

\usepackage{nicefrac}

\usepackage{booktabs}

\usepackage{multirow}
\usepackage{subcaption}
\usepackage{caption}

\usepackage{threeparttable}
\usepackage{siunitx}

\usepackage{makecell}
\usepackage{flushend}

\begin{document}
\title{Learning Families of Fixed-Wing Aircraft Control Policies with Hypernetwork-Conditioned Reinforcement Learning}

\author{Dennis~J.~Marquis and Mazen~Farhood
\thanks{D. Marquis and M. Farhood are with the Kevin T. Crofton Department of Aerospace and Ocean Engineering, Virginia Tech, Blacksburg, VA 24061, USA (e-mail: \{dennisjm, farhood\}@vt.edu).}}

\maketitle

\begin{abstract}
This paper investigates whether it is more effective to condition reinforcement learning policies across changing dynamics using standard input augmentation or hypernetwork-based parameter modulation. We use a small fixed-wing uncrewed aircraft under actuator failures as a demanding test case, as these failures introduce distinctly different closed-loop dynamics. We evaluate parameter-efficient hypernetwork formulations based on Feature-wise Linear Modulation (FiLM) and Low-Rank Adaptation (LoRA), trained using proximal policy optimization. High-fidelity six-degree-of-freedom simulations demonstrate that hypernetwork-conditioned policies improve robustness compared to standard multilayer perceptron policies trained with augmented inputs. Most notably, the hypernetwork policies generalize effectively to time-varying actuator failure modes not encountered during training, indicating that modulating internal parameters provides significant advantages over standard input augmentation for this class of highly coupled dynamics.
\end{abstract}

\IEEEpeerreviewmaketitle

\section{Introduction}
\label{sec:intro}

Reinforcement learning (RL) is a flexible framework for designing control policies for complex systems, enabling controllers to be learned through interaction with nonlinear, uncertain environments. For small uncrewed aircraft systems (sUAS), RL has been applied across a range of tasks from low-level stabilization to high-level planning~\cite{Azar2021,Kurunathan2024}, but its deployment in real-world settings remains challenging. In particular, achieving robustness to changing system conditions is critical, as variations in dynamics may arise from actuator degradation, environmental disturbances such as wind, modeling errors, or operation outside the nominal flight envelope. RL policies are often implemented as multilayer perceptrons (MLPs), which are feedforward neural networks composed of stacked linear layers and nonlinear activation functions. While expressive, such architectures must represent behavior across all operating conditions using a single set of parameters, and performance can degrade when the system deviates from the conditions encountered during training.

A key challenge in learning policies that must operate across a range of system conditions is gradient interference, where updates from different operating regimes can conflict and degrade learning performance~\cite{Sener2018,Yu2020}. In standard MLP policies, data collected under different dynamics lead to updates that can push shared parameters in opposing directions, leading to overly conservative solutions, overfitting to dominant regimes, or unstable training. This issue is particularly pronounced when changes in the environment induce structural changes in the system dynamics. Actuator failures in fixed-wing sUAS represent one such case. For example, a rudder failure reduces yaw authority and modifies the coupling between yaw, roll, and lateral velocity dynamics. The policy must therefore capture multiple qualitatively different control strategies for different failures. One possible solution is to use a set of MLP controllers with switching logic based on the operating condition. However, switched MLPs require discretizing the failure space, which creates too many modes as the system's dimensionality increases.

An alternative way to address the gradient interference issue is to use hypernetworks~\cite{Ha2017}, which map a set of scheduling variables to the parameters of a main network, enabling the representation of a family of specialized controllers rather than a single static policy. This allows the policy to vary with the scheduling variables. Recent surveys~\cite{Chauhan2024} highlight the growing use of hypernetworks across a range of domains. In multi-objective optimization settings~\cite{Navon2021}, a hypernetwork generates a family of policies in which preference inputs lead to different Pareto optimal solutions. In RL, hypernetworks have been used to estimate gradients of policies or value functions~\cite{Sarafian2021}. Hypernetworks have also been incorporated into proximal policy optimization (PPO) \cite{Schopf2022}.

In parallel, recent advances in large-scale machine learning have focused on leveraging hypernetworks for adapting (``fine-tuning'') pretrained models, where the goal is to modify model behavior for new tasks without retraining from scratch~\cite{Oswald2020}. Parameter-efficient methods such as Feature-wise Linear Modulation (FiLM)~\cite{Perez2018} and Low-Rank Adaptation (LoRA)~\cite{Hu2022} enable adaptation without requiring the hypernetwork to generate all of the main-network weights, instead modulating the main network using a small number of parameters. Notably, FiLM has been shown to improve generalization to challenging and previously unseen inputs, including zero-shot reasoning tasks on the CLEVR dataset. These approaches have been adopted in large-scale models, particularly in natural language processing, but their use as a mechanism for control policy conditioning in RL remains relatively unexplored.

Herein, we investigate hypernetwork-conditioned policies for robust control of fixed-wing sUAS under parameterized actuator faults, where the actuator-condition vector is available to the controller, e.g., from actuator-position sensing. Both the hypernetwork and base policy are trained jointly end-to-end using PPO. To avoid the computational cost of generating full network weights, we use parameter-efficient formulations based on FiLM and LoRA. Our objective is not to propose a new fault-tolerant control architecture or benchmark against domain-specific methods \cite{VerhaegenEtAl2010,LuEtAl2016,GavraVanKampen2024,Giral2026}. Instead, by evaluating these parameterized failures on a six-degree-of-freedom sUAS, we establish a sufficiently complex family of dynamics to serve as a suitable testbed for a key representation question: is it more effective to append scheduling variables to a fixed-weight policy, or to use them to modulate the policy's internal parameters? All policies share the same base architecture and training setup to compare these conditioning mechanisms.

The contributions of this work are as follows. First, we formulate fixed-wing aircraft control across actuator-dependent dynamics as a conditional-policy learning problem. Second, by comparing FiLM and LoRA against direct input augmentation, we demonstrate that internal policy modulation significantly improves generalization to unseen and time-varying failure profiles, particularly for strongly coupled rudder failures. Last, we give design guidance for conditional neural controllers by analyzing the effects of critic conditioning, LoRA rank, policy sensitivity, and~inference~cost.

This paper is organized as follows. Section~\ref{sec:prelim} introduces key preliminaries, including hypernetwork-based policy adaptation and the fixed-wing sUAS model. Section~\ref{sec:framework} presents the learning framework, describing the simulation environment, action and observation spaces, reward design, and reference path generation. Section~\ref{sec:setup} details the experimental setup, including the controller architectures, training configuration, and evaluation protocol. Section~\ref{sec:results} presents the results, including studies of the FiLM and LoRA hypernetwork architectures, sensitivity observations, computational considerations, and example simulations. Finally, Section~\ref{sec:conclude} concludes the paper.

\section{Preliminaries}
\label{sec:prelim}

\subsection{Hypernetwork-Based Policy Adaptation}
Hypernetworks adapt a policy by mapping a conditioning variable to the parameters of a main network. A standard approach is to generate the full set of weights and biases, but this can be computationally expensive, thus motivating parameter-efficient alternatives. We consider FiLM and LoRA, which are used for adapting large language models.

Let $\mathbf{o}_k$ denote the observation and $\mathbf{a}_k$ the resulting action at instant $k$. The main network is a feedforward network with $L$ layers. Let $\mathbf{h}_k^{(\ell)}$ denote the hidden state at layer $\ell$, with $\mathbf{h}_k^{(0)} = \mathbf{o}_k$. The main (unadapted) network is defined as
\begin{equation*}
\mathbf{h}_k^{(\ell+1)} = \sigma\!\big( W^{(\ell)} \mathbf{h}_k^{(\ell)} + \mathbf{b}^{(\ell)} \big), \quad \ell = 0,\dots,L-1,
\end{equation*}
where $W^{(\ell)}$ and $\mathbf{b}^{(\ell)}$ denote the weight matrix and bias vector of layer $\ell$, and $\sigma(\cdot)$ is a nonlinear activation function (here, $\tanh$). The output layer produces the action $\mathbf{a}_k = \mathbf{h}_k^{(L)}$.

\vspace{0.5em}
\noindent\textbf{FiLM:}
FiLM adapts the main network by applying feature-wise affine transformations to intermediate activations,
\begin{equation*}
\mathbf{h}_k^{(\ell+1)} = \sigma\!\big( \mathbf{p}_\mathrm{scale}^{(\ell)} \odot (W^{(\ell)} \mathbf{h}_k^{(\ell)} + \mathbf{b}^{(\ell)}) + \mathbf{p}_\mathrm{shift}^{(\ell)} \big),
\end{equation*}
where $\mathbf{p}_\mathrm{scale}^{(\ell)}$ and $\mathbf{p}_\mathrm{shift}^{(\ell)}$ denote element-wise scaling and shifting vectors generated by the hypernetwork, and $\odot$ denotes element-wise multiplication.

\vspace{0.5em}
\noindent\textbf{LoRA:}
LoRA adapts the main network by introducing low-rank updates to the weight matrices,
\begin{equation*}
W^{(\ell)} \;\rightarrow\; W^{(\ell)} + U^{(\ell)} \operatorname{diag}(\mathbf{r}^{(\ell)}) V^{(\ell)T},
\end{equation*}
where $U^{(\ell)} \in \mathbb{R}^{n_{\ell+1} \times n_r}$, $V^{(\ell)} \in \mathbb{R}^{n_{\ell} \times n_r}$ are learned matrices,  $\mathbf{r}^{(\ell)} \in \mathbb{R}^{n_r}$ is generated by the hypernetwork, and $\operatorname{diag}(\mathbf{r}^{(\ell)})$ denotes the diagonal augmentation of the elements of $\mathbf{r}^{(\ell)}$. The rank $n_r$ controls the expressiveness of the adaptation.

Both approaches greatly reduce the dimensionality of the hypernetwork output. In this implementation, the hypernetwork takes a vector describing actuator failures as input and generates the corresponding adaptation parameters. The hypernetwork and main network parameters are trained jointly using PPO, in contrast to large language model settings where adaptation is typically applied to a pre-trained network. FiLM modulation is applied to the hidden layers via $\mathbf{p}_{\mathrm{scale}}^{(\ell)} = \mathbf{1} + 0.1 \mathbf{p}_{\mathrm{h,scale}}^{(\ell)}$ and $\mathbf{p}_{\mathrm{shift}}^{(\ell)} = 0.1 \mathbf{p}_{\mathrm{h,shift}}^{(\ell)}$, where $\mathbf{p}_{\mathrm{h,scale}}$ and $\mathbf{p}_{\mathrm{h,shift}}$ denote the hypernetwork outputs. Similarly, the LoRA hypernetwork output is scaled by a factor of $1/n_r$. This parameterization serves a similar purpose to the principled initialization strategies in~\cite{Chang2020}.

\subsection{sUAS Dynamics}
\label{sec:dynamics}

This section presents the six-degree-of-freedom nonlinear model of a fixed-wing sUAS based on the CZ-150 platform. The vehicle position in the inertial North–East–Down frame $\mathcal{F}_I$ is denoted by $\mathbf{p}=[x\;y\;z]^T$. Linear and angular velocities expressed in the body-fixed frame $\mathcal{F}_b$ are $\mathbf{v}=[u\;v\;w]^T$ and $\boldsymbol{\omega}=[p\;q\;r]^T$, respectively. The vehicle attitude is parameterized by the Euler angles $\boldsymbol{\theta}=[\phi\;\theta\;\psi]^T$, which define the rotation matrix $R_{Ib}(\boldsymbol{\theta})$ from $\mathcal{F}_b$ to $\mathcal{F}_I$. From the inertial velocity $\dot{\mathbf p}$, the flight path angle $\gamma = \arctan2(-\dot{z},\sqrt{\dot{x}^2+\dot{y}^2})$ and the course angle $ \chi = \arctan2(\dot{y},\dot{x})$ can be derived.

Assuming symmetry about the $x$–$z$ plane, the inertia matrix contains the nonzero elements $J_{xx},J_{yy},J_{zz}$ and $J_{xz}$. Let $m$ denote the vehicle mass, $g$ the gravitational constant, and $\bar{c}$, $b$, and $S$ the mean aerodynamic chord, wingspan, and reference wing area. Wind disturbances are represented by the inertial-frame velocity $\mathbf{v}_w=[v_{w,x}\;v_{w,y}\;v_{w,z}]^T$. The air-relative velocity in $\mathcal{F}_b$ is $\mathbf{v}_r=[u_r\;v_r\;w_r]^T=\mathbf{v}-R_{Ib}^T(\boldsymbol{\theta})\mathbf{v}_w.$

The commanded inputs $\boldsymbol{\delta}^{\mathrm{cmd}} = [\delta_E^{\mathrm{cmd}} \; \delta_A^{\mathrm{cmd}} \; \delta_R^{\mathrm{cmd}} \; \delta_T^{\mathrm{cmd}}]^T$ are propagated through first-order actuator dynamics to produce the actuator outputs $\boldsymbol{\delta} =
[\delta_E \; \delta_A \; \delta_R \; \delta_T]^T$, corresponding to elevator, aileron, rudder, and throttle. Surface deflections are measured in radians and the throttle command in revolutions per second. The resulting equations of motion are
\begin{equation*}
\begin{aligned}
\dot{\mathbf{p}} &= R_{Ib}(\boldsymbol{\theta})\mathbf{v}, \quad
\dot{\boldsymbol{\theta}} = \varepsilon(\phi,\theta)\boldsymbol{\omega}, \\
\dot{\mathbf{v}} &= \mathbf{v}\times\boldsymbol{\omega}
+R_{Ib}^T(\boldsymbol{\theta})[0\;0\;g]^T
+\frac{1}{m}\mathbf{F}(\mathbf{v}_r,\boldsymbol{\omega},\boldsymbol{\delta}), \\
\dot{\boldsymbol{\omega}} &= J^{-1}\!\left(J\boldsymbol{\omega}\times\boldsymbol{\omega}+\mathbf{M}(\mathbf{v}_r,\boldsymbol{\omega},\boldsymbol{\delta})\right),
\end{aligned}
\end{equation*}
where $\varepsilon(\phi,\theta)$ is the standard Euler rate matrix.
The aerodynamic forces and moments are parameterized as
\[
\mathbf{F} = \bar{q}S[C_X \quad C_Y \quad C_Z]^T, \
\mathbf{M} = \bar{q}S[C_Lb \quad C_M\bar{c} \quad C_Nb]^T,
\]
where $C_i$ for $i \in \{X, Y, Z\}$ and $i \in \{L, M, N\}$ represent the non-dimensional aerodynamic force and moment coefficients, respectively, and the arguments of the functions $\mathbf{F}$, $\mathbf{M}$, and $C_i$ are suppressed for simplicity. Here, the dynamic pressure is $\bar q=\tfrac{1}{2}\rho_{\text{air}}V_r^2$, where $V_r=\|\mathbf{v}_r\|$ is the airspeed and $\norm{\cdot}$ denotes the Euclidean norm.

The aerodynamic regressors include the angle of attack $\alpha=\arctan(w_r/u_r)$, sideslip angle $\beta=\arcsin(v_r/V_r)$, and non-dimensional angular rates $\hat p = {pb}/{2V_r}$, $\hat q = {q\bar c}/{2V_r}$, $\hat r = {rb}/{2V_r}.$ Additional regressors include the elevator deflection $\delta_E$, rudder deflection $\delta_R$, and the inverse advance ratio $\mathcal{J} = {\delta_T D_{\mathrm{prop}}}/{V_r},$ where $D_{\mathrm{prop}}$ denotes the propeller diameter.

On the physical CZ-150 platform, a single aileron command drives both the left and right aileron servos through a shared control signal. The two surfaces therefore move symmetrically during nominal operation. However, when modeling actuator faults, the physical deflections of the left and right ailerons may differ. To capture this effect, we introduce the individual surface deflections $\delta_{A_l}$ and $\delta_{A_r}$. The aerodynamic model depends on the effective aileron deflection $\delta_A^{\mathrm{eff}} = \tfrac{1}{2}(\delta_{A_l} + \delta_{A_r})$, which reduces to $\delta_A$ when the two surfaces move symmetrically. In addition, asymmetric aileron deflections produce a pitching moment. This effect is captured through an additional term in $C_M$ proportional to $\delta_A^{\mathrm{diff}}=\delta_{A_r}-\delta_{A_l}$.

The aerodynamic model retains the structure we identified experimentally in~\cite{Marquis2026}, with $\delta_A$ replaced by $\delta_A^{\mathrm{eff}}$ where appropriate and the additional asymmetric pitching term included in $C_M$, with coefficient $C_{M_{\delta_A^{\mathrm{diff}}}}=-0.02$. The aerodynamic, inertial, geometric, and actuator parameters are otherwise identical to those in~\cite{Marquis2026}.
To enable path following, the virtual vehicle formulation of~\cite{Fry2020} is adopted, where the aircraft states are expressed relative to a reference vehicle moving along the desired path.

\section{Learning Framework}
\label{sec:framework}

\subsection{Simulation Environment}
\label{sec:sim}
Training is conducted in a custom Python simulation environment based on the CZ-150 aircraft dynamics described in Section~\ref{sec:dynamics}. The environment follows the structure introduced in our prior work~\cite{Marquis2026b}, with several sources of stochasticity included to better approximate real flight conditions. Gaussian measurement noise is applied to all sensor channels, with standard deviations of $0.01$\,rad/s for $\boldsymbol{\omega}$, $2$\,m/s for $V_r$, $0.01$\,rad for $\phi$ and $\theta$, $0.1$\,rad for $\psi$, $0.02$\,rad for $\chi$, $0.03$\,m for $x$ and $y$, $0.01$\,m for $z$, and $0.03$\,m/s$^2$ for the translational acceleration $\mathbf{f}=[f_x\;f_y\;f_z]^T$. Wind is modeled as a steady component with randomly sampled magnitude in $[3,5]$\,m/s and heading in $[0^\circ,360^\circ]$, combined with gust disturbances generated using the Dryden turbulence model~\cite{Real1993} with a moderate reference wind speed of 30~knots at low altitude. Controller inputs are delayed by up to one simulation step. To model aerodynamic uncertainty, each coefficient $C_{i,k}$ for $i \in \{X,Y,Z,L,M,N\}$ is perturbed as
\begin{equation*}
C_{i,k} \rightarrow C_{i,k} + \Delta_{C_i,k},
\end{equation*}
where $\boldsymbol{\Delta}_{\mathbf{C},k}\in\mathbb{R}^6$ satisfies the same magnitude and rate bounds defined in~\cite{Marquis2026b}. Unlike the adversarial training setup used in that work, the perturbations here are sampled stochastically within these bounds. The continuous-time dynamics are integrated using a fourth-order Runge--Kutta scheme with fixed step size $\Delta t = 0.04$\,s, yielding the discrete-time sequence $x_k = x(t_k)$ with $t_k = k\Delta t$.

\subsection{Failure Parameterization}
\label{sec:failure}
Actuator faults are modeled as a loss of control authority through a fixed (stuck) deflection. In this work, the possible failure modes involve the right aileron, left aileron, and rudder. Each actuator is associated with (i)  a binary failure parameter, $\lambda_{i,k}^{\mathrm{fail}}$, that indicates whether the actuator is stuck or operating normally, and (ii) a stuck deflection level, $\lambda_{i,k}^{\mathrm{val}}$, expressed as a fraction of the actuator's upper saturation limit $\delta_i^{\mathrm{sat}}$ for $i=A_r,A_l,R$ (as is typically the case for fixed-wing aircraft, the lower saturation limit is assumed to be $-\delta_i^{\mathrm{sat}}$).  The failure parameters are collected in a vector $\boldsymbol{\lambda}_k \in \mathbb{R}^6$:
\begin{equation*}
\boldsymbol{\lambda}_k = [
\lambda_{A_r,k}^{\mathrm{fail}}, \lambda_{A_r,k}^{\mathrm{val}}, \;
\lambda_{A_l,k}^{\mathrm{fail}}, \lambda_{A_l,k}^{\mathrm{val}}, \;
\lambda_{R,k}^{\mathrm{fail}}, \lambda_{R,k}^{\mathrm{val}}
]^T.
\end{equation*}
For each actuator, the stuck deflection is given by $\delta_{i,k}^{\mathrm{stuck}} = \lambda_{i,k}^{\mathrm{val}} \, \delta_i^{\mathrm{sat}}$, where $\lambda_{i,k}^{\mathrm{val}} \in [-1,1]$. The binary parameter $\lambda_{i,k}^{\mathrm{fail}} \in \{0,1\}$, with  a value of zero indicating nominal operation and a value of one corresponding to stuck behavior.

This formulation enables a wide range of actuator degradations between nominal operation and fully stuck behavior, while allowing the stuck deflection to vary over the actuator limits. In this work, we assume the fault descriptor $\boldsymbol{\lambda}_k$ is available during execution, e.g., from a fault-diagnosis module using commanded and encoder-measured actuator positions. Alternatively, this requirement can be relaxed by introducing an estimator that infers each $\boldsymbol{\lambda}_k$ from the vehicle's state and control history, e.g., by modifying the approach in~\cite{Guo2023}, enabling operation without direct actuator sensing.

\subsection{Action and Observation Spaces}
\label{sec:spaces}
For each policy, the actions are control perturbations relative to the reference input (i.e., the trim input associated with the reference path), given by
$\mathbf{a}_k = \boldsymbol{\delta}_k^{\mathrm{cmd}} - \boldsymbol{\delta}_{\mathrm{ref},k}^{\mathrm{cmd}}$.

The available measurements on the CZ-150 are angular velocity, airspeed, attitude, inertial position, and translational acceleration. The measurement vector is defined as $\mathbf{y}_k = [\boldsymbol{\omega}_k^T \;\; V_{a,k} \;\; \boldsymbol{\theta}_k^T \;\; \mathbf{p}_k^T \;\; \mathbf{f}_k^T]^T$,
and the tracking error is $\bar{\mathbf{y}}_k = \mathbf{y}_k - \mathbf{y}_{\mathrm{ref},k},$ where the position component of $\bar{\mathbf{y}}_k$ is expressed in the body frame via $R_{Ib}(\boldsymbol{\theta}_k)^T(\mathbf{p}_k - \mathbf{p}_{\mathrm{ref},k})$. The observation vector $\mathbf{o}_k$ is constructed to capture tracking performance, control context, and geometric information relevant for path following. In addition to $\bar{\mathbf y}_k$, the observation vector includes the reference input $\boldsymbol{\delta}_{\mathrm{ref},k}^{\mathrm{cmd}}$, the previous control input $\boldsymbol{\delta}_{k-1}^{\mathrm{cmd}}$, the control margin $\mathbf{m}_k$, the maneuver parameters $\kappa_k$ and $\gamma_k$ (see \S \ref{sec:ref}), inertial-frame position error $\bar{\mathbf p}_k = \mathbf p_k - \mathbf p_{\mathrm{ref},k}$, course angle representation $\boldsymbol{\chi}_k = [\sin(\chi_k) \; \cos(\chi_k)]^T$, and course angle error $\boldsymbol{\chi}_{\mathrm{err,k}} = [\sin(\chi_k - \chi_{\mathrm{ref},k}) \; \cos(\chi_k - \chi_{\mathrm{ref},k})]^T$. The control margin $\mathbf{m}_k$ quantifies proximity to actuator saturation:
\begin{equation*}
\mathbf{m}_k = \min \Bigg\{
\max\Big(\mathbf{0}, \tfrac{\boldsymbol{\delta}^{\mathrm{sat}} - \boldsymbol{\delta}_k^{\mathrm{cmd}}}{\boldsymbol{\delta}^{\mathrm{sat}} - \boldsymbol{\delta}_{\mathrm{ref},k}^{\mathrm{cmd}}}\Big),
\max\Big(\mathbf{0}, \tfrac{\boldsymbol{\delta}_k^{\mathrm{cmd}} + \boldsymbol{\delta}^{\mathrm{sat}}}{\boldsymbol{\delta}_{\mathrm{ref},k}^{\mathrm{cmd}} + \boldsymbol{\delta}^{\mathrm{sat}}}\Big)
\Bigg\},
\end{equation*}
where all operations in the above equation (e.g., division, $\max$, $\min$, etc.) are applied element-wise. By construction, $\mathbf{m}_k = \mathbf{1}$ at trim and approaches zero as saturation is reached.

Compared to~\cite{Marquis2026b}, the inclusion of both inertial-frame and body-frame position errors gives additional geometric context that improves path-following performance. The incorporation of course angle information further enhances tracking. A trigonometric parameterization is used to avoid discontinuities associated with angle wrapping. For the MLP policy, the failure parameter vector $\boldsymbol{\lambda}_k$ is included in the observation, increasing the observation vector dimension from 34 to 40. For hypernetwork-based policies, $\boldsymbol{\lambda}_k$ is used to condition the network weights and so is not appended to the main network's observation. All observations are normalized~to~$[-1,1]$.

\subsection{Reward Design}
The reward at each instant $k$ is composed of tracking and control terms,
$R_k = R_{\mathrm{tracking},k} + R_{\mathrm{input},k}$. The tracking reward is shaped using a set of normalized exponentials,
\begin{equation*}
R_{\mathrm{tracking},k} = \sum_{i=1}^{9} k_{1,i} \exp\!\big(-k_{2,i} |\bar{y}_{k,i}|\big),
\end{equation*}
where $\bar{y}_{k,i}$ denotes the $i$-th component of the tracking error. The indices $i=1,2,3$ correspond to body rates $(p,q,r)$ with $k_{1,i}=0.1$, $k_{2,i}=1.0$; $i=4,5,6$ correspond to $(\phi,\theta,\chi)$ with $k_{1,i}=0.2$, $k_{2,i}=5.0$; and $i=7,8,9$ correspond to $(x,y,z)$ with $k_{1,i}=0.5$, $k_{2,i}=0.37$. The weights $k_{1,i}$ and $k_{2,i}$ are selected based on allowable deviations from trim. Position tracking receives the highest weight as geometric path following is the primary objective. Angular rate tracking is included to promote training stability. The course angle $\chi$ is used in place of yaw $\psi$ to define the heading-related tracking error. Under wind disturbances, effective tracking may require sustained sideslip (``crabbing''), which would be unnecessarily penalized if using yaw. The course angle measures alignment of the velocity vector with the path direction, making it a more appropriate metric for path-following objectives.

The input reward penalizes both proximity to actuator saturation and rapid changes in control inputs,
\begin{equation*}
R_{\mathrm{input},k}
= k_3 \sum_{j=1}^{4} \log\!\big(m_{k,j} + 10^{-6}\big)
- k_4 \left\lVert \boldsymbol{\delta}_k^{\mathrm{cmd}} - \boldsymbol{\delta}_{k-1}^{\mathrm{cmd}} \right\rVert^2,
\end{equation*}
where $k_3=0.02$ and $k_4=0.2$. The first term acts as a barrier function that discourages operation near actuator limits, while the second term regularizes control rates to avoid high-frequency or bang-bang behavior.

\begin{rmk} Sparse reward formulations based on banded error thresholds were also investigated. For each error component, a reward of 1 is assigned if the error lies within a tight tolerance, 0.3 if within a looser tolerance, and 0 otherwise. These banded rewards are applied to position, angular rate, and attitude errors. While this formulation simplifies the reward structure, it generally degrades performance for most policies. However, the FiLM-conditioned policies exhibited negligible performance degradation due to the sparse reward, suggesting a reduced reliance on dense reward shaping.
\end{rmk}

\subsection{Reference Path Generation}
\label{sec:ref}
Reference paths are generated by concatenating motion primitives following the approach in~\cite{Arifianto2015}. The construction closely follows that of~\cite{Marquis2026b}, where straight-and-level and coordinated turning segments are combined to form lemniscate-like paths. Each segment is associated with a steady-state trim defined by equilibrium states and inputs, computed at a nominal airspeed of $V_{r,\mathrm{nom}} = 21\,\mathrm{m/s}$ using a nonlinear solver. The vehicle state is defined as $\mathbf{x} = [\mathbf{p}^T \; \mathbf{v}^T \; \boldsymbol{\theta}^T \; \boldsymbol{\omega}^T \; \boldsymbol{\delta}^T \; \chi]^T$. In the absence of wind, the course angle coincides with the yaw angle, i.e., $\chi = \psi$. For each trim condition, time-varying reference histories are generated by integrating the vehicle dynamics, producing references for $x$, $y$, $z$, $\psi$, and $\chi$. The resulting segments are concatenated to obtain a complete reference trajectory $\mathbf{x}_{\mathrm{ref}}$ and input $\boldsymbol{\delta}_{\mathrm{ref}}^{\mathrm{cmd}}$. Coordinated turns are parameterized by the inverse radius of curvature $\kappa$ and flight path angle $\gamma$, selected from the sets
\begin{equation*}
\begin{aligned}
K_{\mathrm{ref}} &= \{-0.02, -0.012, 0.012, 0.02\}, \\
\Gamma_{\mathrm{ref}} &= \{-0.21, -0.11, 0, 0.11, 0.21\},
\end{aligned}
\end{equation*}
with $\kappa = \gamma = 0$ corresponding to straight-and-level flight.

Under actuator failures and wind disturbances, enforcing timing constraints can render trajectory tracking infeasible, as the vehicle may need to deviate from its nominal airspeed to maintain stability. The path-following formulation relaxes these timing constraints, allowing the controller to prioritize geometric tracking of the path.

\section{Experimental Setup}
\label{sec:setup}

\subsection{Controller Architectures}
We consider three policy architectures: a baseline MLP and hypernetwork-conditioned policies using FiLM and LoRA.

\vspace{0.5em}
\noindent\textbf{MLP:}
The baseline policy is a feedforward neural network with two hidden layers of 64 neurons each and $\tanh$ activations. The value function is represented by a network of identical structure. For the MLP, the failure vector $\boldsymbol{\lambda}_k$ is provided as part of the observation vector $\mathbf{o}_k$.

\vspace{0.5em}
\noindent\textbf{FiLM and LoRA:}
These policies consist of a main network with the same architecture as the MLP and a hypernetwork that adapts the main network based on the failure vector $\boldsymbol{\lambda}_k$. The hypernetwork is a feedforward network with two hidden layers of 32 neurons each and $\tanh$ activations. For both FiLM- and LoRA-conditioned policies, the hypernetworks generate parameters that adapt the hidden layers of the main network. For LoRA, multiple ranks $n_r \in \{8,16,32, 48, 64\}$ are evaluated to study the effect of the adaptation capacity.

We also investigate applying hypernetwork-based adaptation to the value function in PPO. Specifically, we compare configurations where the value network is either a standard MLP or shares the same hypernetwork-based conditioning as the policy. We refer to hypernetwork-conditioned FiLM policies simply as FiLM and LoRA-based policies as LoRA~($n_r$), where $n_r$ denotes the adaptation rank. We add +HC (hyper-conditioned critic)  to indicate configurations in which the value network is also conditioned by the hypernetwork.

\subsection{Training Configuration}

All policies are trained using PPO on Virginia Tech's Advanced Research Computing cluster. We use Stable-Baselines3~\cite{Raffin2021StableBaselines3} with PyTorch~\cite{paszke2019pytorch}, and implement a custom hypernetwork-based policy compatible with the SB3 framework. Each policy update is based on rollouts collected from multiple parallel simulation environments. During training, episodes are sampled from a mixture of nominal and failure scenarios. Specifically, we consider (i) nominal episodes with no actuator failures, (ii) episodes where a single actuator is fully stuck for the entire duration, and (iii) episodes where a single actuator becomes fully stuck at a random time during the episode. After the onset instant $N_\mathrm{fail}$, the actuator transitions to a persistent stuck configuration, corresponding to $\lambda_{i,k}^{\mathrm{fail}} = 1$ with fixed $\lambda_{i,k}^{\mathrm{val}}$ for all $k\geq N_\mathrm{fail}$.

The failure parameters are sampled over a discrete grid for the stuck deflection values; for each actuator, $\lambda_{i,k}^{\mathrm{val}}$ is drawn from $\Lambda_\mathrm{train} \coloneqq \{0, \pm0.25, \pm0.5\}$. This restriction avoids failure configurations that render the path-following task dynamically infeasible. For example, a fully stuck right aileron at saturation may require sustained opposing input from the left aileron, leaving insufficient control authority to reject disturbances such as wind. Training hyperparameters, found to produce stable learning, are consistent with those detailed in \cite{Marquis2026b}.

\subsection{Evaluation Protocol}
Trained policies are evaluated over 1,000 episodes per set to assess both in-distribution robustness and out-of-distribution generalization. Performance is quantified by Mean Path Error (MPE) and Maximum Path Error (MaxPE), representing the average and peak position errors per episode, respectively.

First, we assess robustness by simulating single-actuator failures at random onset times. Actuators are selected uniformly, with stuck deflection values drawn from the extended set $\Lambda_\mathrm{eval} \coloneqq \{0, \pm 0.125, \pm 0.25, \pm 0.375, \pm 0.5\}$. This grid includes intermediate values to evaluate the policy's ability to interpolate across the failure space.

To further assess generalization to out-of-distribution scenarios, we introduce time-varying actuator faults (``flutter'') not encountered during training. In this setting, a single actuator is subjected to a temporary, nonstationary oscillatory failure. Each flutter episode is defined by a randomly selected actuator, onset instant $N_{\mathrm{fail}}$, and a finite duration of 1\,s to 10\,s. The failure is centered about a deflection value $\bar{\lambda}_i^{\mathrm{val}} \in \Lambda_{\mathrm{eval}}$, where the actual deflection $\lambda_{i,k}^{\mathrm{val}}$ varies in a piecewise-constant manner within the bounded interval $[\bar{\lambda}_i^{\mathrm{val}} - 0.2, \bar{\lambda}_i^{\mathrm{val}} + 0.2]$. With each value held for short durations of 0.2\,s to 1.0\,s, this creates a dynamic loss of control authority that differs fundamentally from the static failures seen during training.
\section{Results}
\label{sec:results}
The top performing policy for each architecture is evaluated against fixed in-distribution failure modes, with interpolated deviations, hereafter referred to as static failures, and time-varying out-of-distribution (flutter) failures, with simulation set averages summarized in Table~\ref{table:best_policies}. Under static conditions, all controllers maintain stability; however, the largest MaxPE across episodes, i.e., the observed worst-case (WC) error, reveals a disparity. While the hypernetwork-conditioned policies exhibit maximum observed static rudder errors of approximately 22\,m, the MLP experiences a peak error of 36.83\,m. A slight performance asymmetry between left and right aileron failures is observed across all policies, likely due to physical asymmetries in the sUAS model, such as propulsion system misalignment. The distinction between architectures is most profound under flutter failures. The hypernetwork policies do degrade slightly but maintain higher consistency with WC errors below 30\,m. In contrast, the baseline MLP shows severe transient tracking degradation, most notably during rudder flutter where the WC error reaches 159.91\,m. While MaxPE standard deviations (SD) are comparable under static failures, the MLP's rudder flutter SD ($\approx$ 30.44\,m) is over five times higher (e.g., $\approx$ 4.84\,m for LoRA), indicating a large variability in flutter scenarios.

\begin{figure*}[t]
    \centering
    \setlength{\abovecaptionskip}{2pt}
    \setlength{\belowcaptionskip}{2pt}

    \begin{subfigure}{0.9\textwidth}
        \centering
        \includegraphics[width=\linewidth]{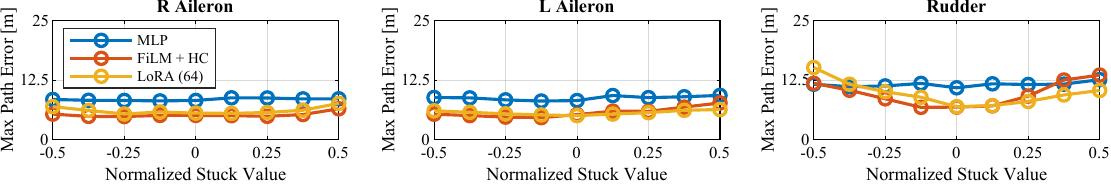}
        \caption{Average MaxPE for static actuator failures}
        \label{fig:random_maxpe}
    \end{subfigure}

    \vspace{2mm}

    \begin{subfigure}{0.9\textwidth}
        \centering
        \includegraphics[width=\linewidth]{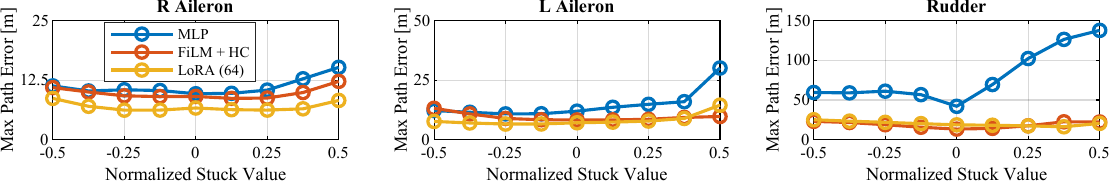}
        \caption{Average MaxPE for flutter failures}
        \label{fig:flutter_maxpe}
    \end{subfigure}

    \caption{Average MaxPE across failure magnitudes for each actuator. Top: static failures. Bottom: flutter failures.}
    \label{fig:maxpe_comparison}\vspace{-4mm}
\end{figure*}

Figure~\ref{fig:maxpe_comparison} presents the average MaxPE values across the range of tested deflection magnitudes. For static failures, all architectures exhibit a nearly uniform and comparable error profile, indicating that the baseline MLP is generally capable of handling these types of failures. However, under flutter failures, the error profiles diverge significantly. While the hypernetwork-conditioned policies maintain average MaxPE values that remain less than the 25\,m early termination threshold used during training, the MLP exhibits a substantial error increase that is most severe for positive rudder deflections. A similar trend occurs for large positive left aileron deflections, where the MLP suffers from much higher error growth than the hypernetwork-conditioned policies. These trends suggest the MLP overfit to the specific dynamics of the static training scenarios, failing to generalize when those same deflections become time-varying. Furthermore, the MLP's higher proficiency in mitigating aileron failures compared to rudder failures suggests a training exposure bias, as aileron failures are encountered twice as frequently as rudder failures due to the inclusion of both left and right aileron failures. The results indicate that conditioning on $\boldsymbol{\lambda}$ enables the hypernetwork policies to better generalize to the new dynamics imposed by the failed actuators.

\begin{table}[h]
\centering
\caption{Performance under static failures and flutter: average MPE, average MaxPE, worst-case (WC) errors. }
\label{table:best_policies}
\resizebox{\columnwidth}{!}{%
\begin{tabular}{ll ccc ccc}
\toprule
 & & \multicolumn{3}{c}{\textbf{Static Failure (m)}} & \multicolumn{3}{c}{\textbf{Flutter (m)}} \\
\cmidrule(lr){3-5} \cmidrule(lr){6-8}
\textbf{Policy} & \textbf{Act.} & \textbf{MPE} & \textbf{MaxPE} & \textbf{WC} & \textbf{MPE} & \textbf{MaxPE} & \textbf{WC} \\
\midrule
\multirow{3}{*}{MLP}
& R Ail. & 3.62 & 8.44 & 20.71 & 4.30 & 11.13 & 32.22 \\
& L Ail. & 3.71 & 8.75 & 19.88 & 5.39 & 14.65 & 34.29 \\
& Rudder & 4.29 & 11.60 & 36.83 & 30.50 & 79.41 & 159.91 \\
\midrule
\multirow{3}{*}{\makecell[l]{FiLM + \\ HC}}
& R Ail. & 1.71 & 5.22 & 18.87 & 4.21 & 9.76 & 20.25 \\
& L Ail. & 1.76 & 5.72 & 18.78 & 3.87 & 9.48 & 23.99 \\
& Rudder & 2.56 & 9.64 & 21.34 & 7.69 & 18.70 & 26.73 \\
\midrule
\multirow{3}{*}{\makecell[l]{LoRA\\ (64)}}
& R Ail. & 2.21 & 6.09 & 18.75 & 2.37 & 6.88 & 20.10 \\
& L Ail. & 2.10 & 5.66 & 19.98 & 3.03 & 8.14 & 19.71 \\
& Rudder & 3.52 & 9.74 & 21.55 & 7.47 & 20.09 & 29.91 \\
\bottomrule
\end{tabular}}\vspace{-2mm}
\end{table}

\subsection{Effectiveness of Hyper-Conditioned Value Functions}

To evaluate whether conditioning the value function improves robustness, we compare FiLM and LoRA architectures with and without a hypernetwork-conditioned critic. Given that rudder failures consistently proved more challenging to mitigate than aileron failures, the results, shown in Table~\ref{table:hc_comparison_rudder}, highlight rudder-specific performance.

For FiLM-based policies, hyper-conditioning the critic significantly improves performance; under static rudder failures, both average MPE and average MaxPE are reduced by 40\% to 50\%. This suggests that for FiLM, modulating the critic's activations is more effective for advantage estimation than treating $\boldsymbol{\lambda}$ as a standard input feature. Conversely, for LoRA-based architectures, the inclusion of HC results in a substantial performance degradation. For LoRA (16), introducing the conditioned critic nearly doubles the error across all metrics. Since both versions have access to the failure state, this degradation suggests that simultaneously adapting both actor and critic weight matrices via low-rank updates introduces larger optimization complexity. This could lead to high-variance gradients or numerical instability during the iterative PPO update, where a standard input-feature representation of $\boldsymbol{\lambda}$ proves more robust for the value function.

\begin{table}[h]
\centering
\caption{Effect of hyperconditioning the critic under rudder failures: average MPE and average MaxPE.}
\label{table:hc_comparison_rudder}
\begin{tabular}{lcccc}
\toprule
 & \multicolumn{2}{c}{\textbf{Static Failure}} & \multicolumn{2}{c}{\textbf{Flutter}} \\
\cmidrule(lr){2-3} \cmidrule(lr){4-5}
\textbf{Policy} & \textbf{MPE} & \textbf{MaxPE} & \textbf{MPE} & \textbf{MaxPE} \\
\midrule
FiLM & 5.33 & 16.49 & 15.72 & 43.67 \\
FiLM + HC & 2.56 & 9.64 & 7.69 & 18.70 \\
LoRA (16) & 4.34 & 14.86 & 11.75 & 31.06 \\
LoRA (16) + HC & 10.40 & 26.68 & 18.53 & 48.66 \\
\bottomrule
\end{tabular}
\end{table}

\subsection{LoRA Rank Sensitivity}
\label{sec:lora}
We evaluate LoRA-based policies across multiple ranks to assess how adaptation capacity influences robustness. Performance metrics for rudder failures, the most critical mode identified, are provided in Table~\ref{table:lora_rank_rudder}. For most static scenarios, performance remains comparable across the tested power-of-two ranks. Sensitivity to rank primarily emerges during flutter failures. As rank increases from $n_r=8$ to $n_r=64$, there is a consistent trend of improved generalization; while the $n_r=8$ policy exhibits significantly higher average MaxPE, the $n_r=64$ variant achieves robustness levels comparable to the FiLM + HC policy.

Notably, this scaling law is not strictly monotonic. An evaluation at $n_r=48$ resulted in significant instability, with MaxPE exceeding 150\,m in the flutter case. While power-of-two ranks are standard for memory alignment and compute efficiency \cite{nvidia_matrix}, this specific outlier highlights that hypernetwork-conditioned RL remains sensitive to subtle architectural and initialization choices. In practice, rank selection acts as a critical tuning parameter, where minor deviations can lead to optimization instability rather than predictable performance gains.

\begin{table}[h]
\centering
\caption{LoRA rank sensitivity under rudder failures: average MPE and average MaxPE.}
\label{table:lora_rank_rudder}
\begin{tabular}{lcccc}
\toprule
 & \multicolumn{2}{c}{\textbf{Static Failure}} & \multicolumn{2}{c}{\textbf{Flutter}} \\
\cmidrule(lr){2-3} \cmidrule(lr){4-5}
\textbf{Policy} & \textbf{MPE} & \textbf{MaxPE} & \textbf{MPE} & \textbf{MaxPE} \\
\midrule
LoRA (8)  & 3.39 & 13.14 & 20.75 & 60.70 \\
LoRA (16) & 4.34 & 14.86 & 11.75 & 31.06 \\
LoRA (32) & 4.24 & 14.03 & 7.09  & 21.70 \\
LoRA (64) & 3.52 & 9.74  & 7.47  & 20.09 \\
\bottomrule
\end{tabular}
\end{table}

\subsection{Hypernetwork Sensitivity Characterization}
We observe a notable relationship between the hypernetwork's Lipschitz bound and the controller performance. The bound ($\mathcal{L}$) limits the sensitivity of the hypernetwork outputs to changes in $\lambda$; a lower value implies a less sensitive parameterization \cite{Szegedy2014}. Following \cite{Scaman2018}, we estimate this bound for each hypernetwork by computing the product of the spectral norms of their respective weight matrices, given that the $\tanh$ activation function is 1-Lipschitz. For the LoRA-based policies, the Lipschitz bound decreases monotonically as the rank $n_r$ increases ($n_r=8,\, \mathcal{L}=28.38$; $n_r=16,\, \mathcal{L}=21.89$; $n_r=32,\, \mathcal{L}=11.16$; $n_r=64,\, \mathcal{L}=3.73$). Lower values of $\mathcal{L}$ correlate with improved tracking performance.

While the FiLM-based architectures lack a tuning parameter equivalent to rank, we investigate the impact of main-network expressivity by applying the hypernetwork to a smaller main network (two layers of 32 neurons). This less expressive configuration exhibited both poor control performance and a higher Lipschitz bound ($\mathcal{L}=18.76$) compared to the standard FiLM + HC policy ($\mathcal{L}=13.67$). Combined, these results motivate the use of spectral normalization to explicitly constrain sensitivity when training future controllers.

\subsection{Computational Considerations}
Table~\ref{table:complexity} summarizes the computational characteristics of select policies. While the MLP architecture takes a 40-dimensional (40D) observation as input, hyper-conditioned policies separate inputs into state (34D) and failure (6D) vectors. Under this framework, the hyper-conditioned policies require less than 35,000 (35k) parameters, an order of magnitude reduction compared to a full hypernetwork generator, which would require roughly 434k parameters to produce all actor/critic weights ($6\, \mbox{(input)} \!\to 32 \to 32 \to\! 13{,}125\, \mbox{(output)}$).

Training time per iteration remains consistent across architectures, indicating that computational cost is dominated by environment simulation and rollout collection rather than network size. Small variations arise from stochastic effects in PPO optimization, such as differences in the states visited or the frequency of early episode terminations. For instance, the increased training time per iteration for FiLM compared to FiLM + HC likely reflects the overhead from more frequent resets and shorter-lived episodes, despite the smaller network.

Deployment costs are similarly negligible; actor forward-pass requirements range from $10^4$ to $10^5$  floating-point operations (FLOPs). Given that low-cost processors (e.g., Raspberry Pi) provide $\sim 10^9$ FLOPs per second, their computational capacity exceeds the requirements of our 25 Hz control rate by several orders of magnitude. Importantly, hyper-conditioning the value function increases training parameters but incurs no additional runtime cost.

\begin{table}[h]
\centering
\caption{Computational characteristics of trained policies.}
\label{table:complexity}
\begin{tabular}{lccc}
\toprule
\textbf{Policy} & \textbf{Parameters} & \textbf{Train Time / Iter} & \textbf{FLOPs} \\
\midrule
MLP & 13,897 & 24.0\,s & 14k \\
FiLM & 23,405 & 28.7\,s & 32k \\
FiLM + HC & 31,510 & 23.8\,s & 32k \\
LoRA (16) & 19,629 & 28.8\,s & 26k \\
LoRA (16) + HC & 23,958 & 33.7\,s & 26k \\
LoRA (32) & 24,301 & 31.2\,s & 37k \\
LoRA (64) & 33,645 & 30.8\,s & 57k \\
\bottomrule
\end{tabular}\vspace{-2.5mm}
\end{table}

\subsection{Simulation Example}
\begin{figure}[ht]
\centering
\setlength{\abovecaptionskip}{2pt}
\setlength{\belowcaptionskip}{2pt}
\includegraphics[scale=1.1]{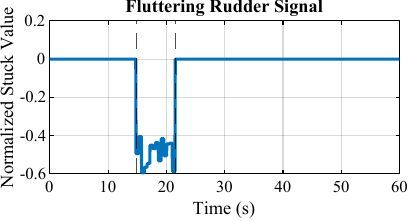}
\caption{Example rudder flutter signal from a simulation episode. This type of conditioning signal is used to assess generalization to unseen, nonstationary failures.}
\label{fig:simulation_flutter}
\end{figure}

\begin{figure*}[t]
\centering
\setlength{\abovecaptionskip}{2pt}
\setlength{\belowcaptionskip}{2pt}
\begin{subfigure}[c]{0.32\textwidth}
    \centering
    \includegraphics[scale=0.78]{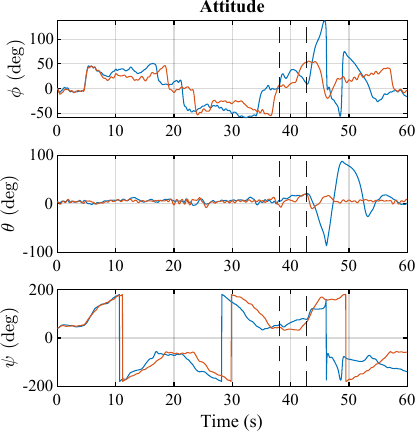}
\end{subfigure}
\hfill
\begin{subfigure}[c]{0.32\textwidth}
    \centering
    \includegraphics[scale=0.78]{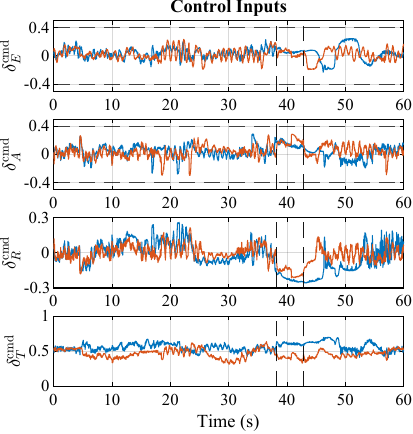}
\end{subfigure}
\hfill
\begin{subfigure}[c]{0.32\textwidth}
    \centering
    \includegraphics[scale=0.78]{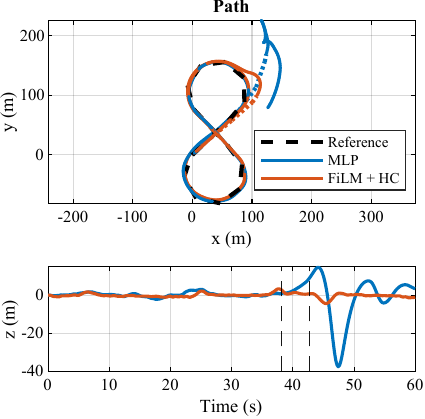}
\end{subfigure}
\caption{State and control histories for an MLP  WC episode under rudder flutter, compared against the FiLM + HC policy.}
\label{fig:simulation_example_mlp}\vspace{-2mm}
\end{figure*}

To illustrate the differences between policies qualitatively, we compare the MLP and FiLM + HC policies across two representative worst-case simulations under rudder flutter. Figure~\ref{fig:simulation_flutter} shows an example stuck value signal, corresponding to approximately 10\,s of flutter. For each simulation, the environmental conditions (e.g., wind, measurement noise, etc.) are identical. In the MLP worst-case scenario (Figure~\ref{fig:simulation_example_mlp}), the policy exhibits excessive bank and pitch angles, leading to a significant altitude gain of nearly 40\,m and substantial path deviation during the flutter episode (indicated by dashed lines). This tracking degradation is exacerbated by the MLP's lack of the aggressive aileron compensations observed in the FiLM + HC policy, which instead utilizes roll-to-yaw coupling to maintain stability.

\begin{figure*}[t]
\centering
\setlength{\abovecaptionskip}{2pt}
\setlength{\belowcaptionskip}{2pt}
\begin{subfigure}[c]{0.32\textwidth}
    \centering
    \includegraphics[scale=0.78]{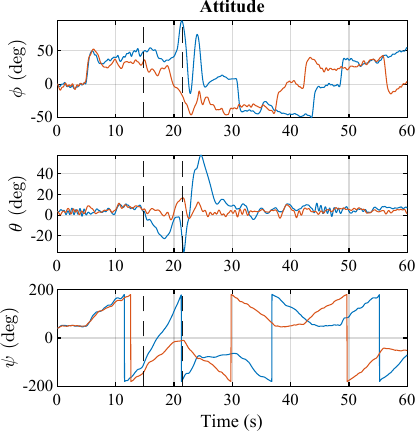}
\end{subfigure}
\hfill
\begin{subfigure}[c]{0.32\textwidth}
    \centering
    \includegraphics[scale=0.78]{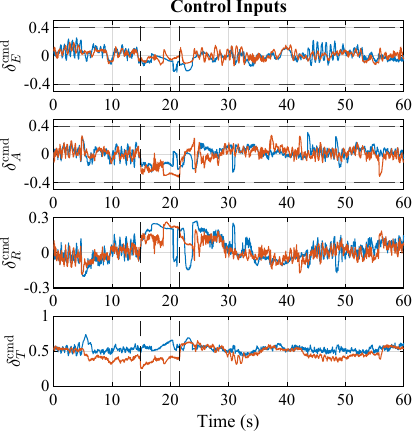}
\end{subfigure}
\hfill
\begin{subfigure}[c]{0.32\textwidth}
    \centering
    \includegraphics[scale=0.78]{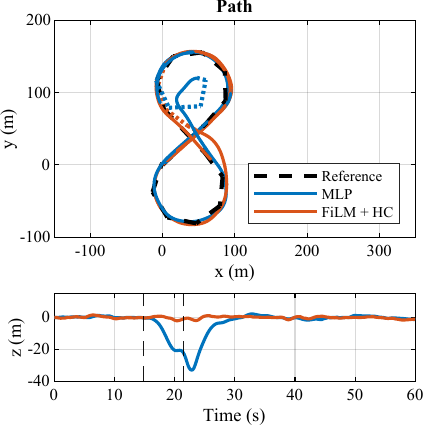}
\end{subfigure}
\caption{State and control histories for a FiLM + HC  WC episode under rudder flutter, compared against the MLP policy.}
\label{fig:simulation_example_film}\vspace{-4mm}
\end{figure*}

In the FiLM + HC worst-case scenario (Figure~\ref{fig:simulation_example_film}), the sUAS deviates from the reference path with a noticeable cross-track error. However, it still maintains tighter overall tracking and altitude control than the MLP. Notably, while the MLP performs worse during the active flutter period, gaining altitude and over-correcting its turn, it demonstrates the ability to effectively recover to the reference path once the rudder returns to nominal operation.

\vspace{-4mm}
\section{Conclusion}\label{sec:conclude}
This work examines how known operating-condition parameters should be incorporated into a reinforcement-learned controller when those parameters index substantially different dynamics. Using fixed-wing actuator failures as a stress test,  FiLM- and LoRA-based hypernetwork modulation are found to generalize more reliably than direct input augmentation, especially to nonstationary failure profiles excluded from training. High-fidelity  six-degree-of-freedom simulations show that these hypernetwork-conditioned policies generalize effectively to demanding failure modes, such as rudder flutter, where the MLP baseline policy diverges. Results indicate that performance is sensitive to architectural choices, including the LoRA rank and the inclusion of a hyper-conditioned value function. In future work, we will incorporate spectral normalization to constrain hypernetwork sensitivity and carry out flight-test experiments to validate performance.

\end{document}